\documentclass[12pt, preprint]{aastex63}
\usepackage{amsmath}
\usepackage{amssymb}
\usepackage{graphicx}
\usepackage{color}
\usepackage{xcolor}
\usepackage{xfrac}
\usepackage[caption=false]{subfig}
\usepackage{rotating}
\usepackage{hyperref}

\newcommand{\be}{\begin{equation}}
\newcommand{\ee}{\end{equation}}       
\newcommand{\bea}{\begin{eqnarray}}
\newcommand{\eea}{\end{eqnarray}}

\begin{document}


\title{Applying Noether's theorem to matter in the Milky Way: 
evidence for external perturbations and non-steady-state effects from 
{\it Gaia} Data Release 2}

\author[0000-0002-6166-5546]{Susan Gardner}
\affiliation{Department of Physics and Astronomy, 
University of Kentucky, Lexington, KY 40506-0055}

\author[0000-0002-9785-914X]{Austin Hinkel}
\affiliation{Department of Physics and Astronomy, 
University of Kentucky, Lexington, KY 40506-0055}

\author[0000-0002-9541-2678]{Brian Yanny}
\affiliation{Fermi National Accelerator Laboratory, Batavia, IL 60510}

\begin{abstract}
We apply Noether's theorem to observations of main-sequence stars 
from the {\it Gaia} Data Release 2 archive to probe
the matter distribution function of the Galaxy. 
That is, we examine the axial symmetry of 
stars at vertical heights $z$,  $0.2 \le |z| \le 3$ kpc, to probe  
the quality of the angular momentum $L_z$ as an integral of motion.
The failure of this symmetry test would speak to a Milky Way, in both its
visible and dark matter, that is not isolated and/or 
not in steady state.
The left-right symmetry-breaking pattern we have observed, north and south, reveals 
both effects, with a measured deviation from symmetry of 
typically 0.5\%.
We show that a prolate form of the gravitational
distortion of the Milky Way by the Large Magellanic Cloud, 
determined from fits to the Orphan stream by \citet{erkal2019total}, 
is compatible with the size and sign of the axial-symmetry-breaking effects 
we have discovered in our sample of up to 14.4 million 
main-sequence stars, speaking to a distortion of an emergent, rather than static, nature. 
\end{abstract}

\section{Introduction}

The stars of isolated galaxies in steady state have 
distribution functions (DFs) that obey the Poisson and collisionless Boltzmann (Vlasov) 
equations and are controlled by particular integrals of motion, as dictated by Jeans theorem \citep{jeans1915, binney2008GD}. 
In axisymmetric galaxies these integrals include the total energy $E$ 
and the component of angular momentum parallel to the symmetry axis $L_z$, and the axially symmetric DFs are also reflection symmetric about the galactic mid-plane 
\citep{an2017reflection}. 
The ubiquity of flat galactic rotation 
curves \citep{sofue2001rotation} are  
commonly interpreted as galaxies 
embedded in a spherical halo of 
dark matter (DM), for which $E$ and $\mathbf{L}$ are also integrals of motion. 
Here we scrutinize how these expectations are borne out in our Galaxy using observations of 
stars from {\it Gaia} Data Release 2 (DR2) \citep{prusti2016gaia, brown2018gaia}. 

In our Galaxy,  
the estimated stellar relaxation time is far longer than the age of the
universe, making the 
neglect of stellar collisions an excellent approximation.
This, in turn, allows us to model the collection of its
stars as a continuous mass distribution, 
as essential to the use of the DF formalism. 
The continuous symmetries of that mass distribution become key probes of its dynamics, 
for Noether's
theorem links the existence of an integral of motion to that of a continuous
symmetry \citep{noether1918,olver2012applications}. 
Thus to test the extent to which
the angular momentum $L_z$ serves as an integral of motion, we can 
study whether the stellar mass distribution is 
axially symmetric. Of course the Galaxy possesses features that 
break axial symmetry, such as spiral arms \citep{chen2019spiral,reid2019trigonometric} or dust \citep{rezaei2018dust3D}, 
so that to test axial symmetry we select regions so as to minimize such effects. 
For example, we avoid the immediate  
Galactic mid-plane ($z=0)$ region, choosing 
stars at vertical heights $z$ with 
$0.2 \le |z| \le 3$ kpc, noting that the 
the dust has a 
vertical scale height $H_d$ of $94 \pm 22$ pc 
at the Sun's 
location
\citep{drimmel2001dust}{\footnote{This result is two times smaller than what one would extract from Table 1 of \citet{drimmel2001dust}, because
we define the vertical scale height from the behavior of the dust density, $\rho_{\rm dust} \sim \exp(-|z|/H_d)$, as $|z|$ grows large.}} --- and the latest three-dimensional 
dust map considers $|z| < 100$ pc \citep{rezaei2018dust3D}. 
With this, and our other selections, we also avoid vertical structure in 
the spiral arms \citep{camargo2015tracing}. 
That our selection 
is also north-south reflection symmetric is important to establishing the origin of the symmetry-breaking effects we find. 
We test axial and reflection symmetry by comparing the number of stars ``left'' and ``right'' of the anti-center line 
at a Galactocentric azimuth of $\phi=180^\circ$, 
 running from the Galactic Center (GC) through the sun, so that
``left (right)'' refers to $\phi > (<) 180^{\circ} $ with 
$|180^{\circ} - \phi|$, and we do this both for stars
in the north ($z>0$) and south ($z<0$). The appearance of axial-symmetry breaking would reveal that the matter in the 
Milky Way (MW) is subject to external and/or time-varying forces. By comparing axial symmetry breaking in the 
north and south 
we can separate time-varying forces from external ones. 
Particularly, we find that the axial symmetry breaking of the north and south combined is much smaller than that of their difference.  
 Since \citet{an2017reflection} have shown that an axially symmetric galaxy in steady state must be north-south symmetric,  
it is the breaking of axial symmetry as a north-south difference that emerges as a predominantly non-steady-state effect.  
Indeed, from our study we discover a correlated left-right, north-south asymmetry in stellar number counts.
We interpret the smaller left-right axial symmetry breaking in the combination of star counts, north and south, as evidence of 
external or non-isolating forces, though that such forces may also be time dependent is not excluded. 

Considerable evidence exists for imperfections 
throughout the Galactic disk. 
The disk is warped and flared in HI gas \citep{levine2006vertical, kalberla2007dark} and 
in stars \citep{alard2000flaring, ferguson2017milky}, with striking evidence for the latter emerging recently 
from three-dimensional maps of samples of 1339 and 2431 Cepheids, respectively \citep{chen2019intuitive,skowron2019three}. 
Rings \citep{newberg2002ghost, morganson2016mapping} and ripples \citep{pricewhelan2015reinterpretation, xu2015rings} 
have been noted, and vertical, wave-like asymmetries have been 
observed in main-sequence stars  
near the Sun's location from the 
Sloan Digital Sky Survey (SDSS) 
\citep{widrow2012galactoseismology, yanny2013stellar, ferguson2017milky} and from {\it Gaia} 
DR2 \citep{bennett2018vertical}. 
Evidence for axial-symmetry breaking of 
out-of-plane main-sequence 
stars in the north with SDSS  has also been observed \citep{ferguson2017milky}. 
Studies of the DF have been greatly enriched by the astrometry of 
{\it Gaia} DR2 
\citep{prusti2016gaia, brown2018gaia}. Notably,  
\citet{antoja2018dynamically} have discovered striking ``snail shell and ridge'' correlations within the 
position and velocity components of the DF that speak to both axially asymmetric and non-steady-state behavior,
and, as they note, is attributable to the existence of the Galactic bar, spiral arms, as well as external perturbations. 

The particular origins of these various effects are not well-established.
Galactic warps have been thought to have a dynamical origin, appearing and disappearing on time scales short
compared to the age of the universe, due to interactions with the halo and its satellites 
\citep{nelson1995damping, shen2006galactic}, though it has 
also been suggested 
that the warp in HI gas is due to the presence of the Large Magellanic Cloud (LMC) 
\citep{weinberg2006magellanic}. 
The vertical asymmetries in the stellar density may be due to an ancient 
impact, possibly by the Sagittarius (Sgr) dwarf galaxy \citep{widrow2012galactoseismology}, 
with support for the impact hypothesis coming not only from theoretical 
investigations \citep{purcell2011sagittarius, gomez2012vertical}, 
but also from an observed vertical wave in mean 
metallicity \citep{an2019asymmetric}, inferred from 
SDSS photometry, with features similar to the observed density wave. 
The novel phase-space structures noted by \citet{antoja2018dynamically} also offer support to the impact hypothesis, as 
such features had been predicted as a consequence \citep{purcell2011sagittarius, gomez2012signatures, d2016excitation, fux2001order}.
Recently, too, the discovery of stars with retrograde motion in the disk has led to determination
of a previously unidentified ancient impact, from {\it Gaia}-Enceladus (or the {\it Gaia}-Sausage) in the inner halo \citep{helmi2018merger,belokurov2018co}. 
Also \citet{koppelman2018one} have noted significant merger debris and streams
in the halo, which are also 
an expected consequence of ancient impacts --- and a stellar stream has been discovered in the solar neighborhood 
as well \citep{necib2019evidence}. 
The velocity ellipsoid \citep{hagen2019tilt} and DM
distribution \citep{posti2019mass} are not spheroidal either, with the evidence favoring a prolate 
matter distribution. Studies of flaring HI gas in the outer galaxy also support a prolate DM 
distribution \citep{banerjee2011progressively}; these authors note that a prolate halo can support
long-lived warps \citep{ideta2000time}, which would help to explain why they are commonly seen  \citep{banerjee2011progressively}. 
It has also been suggested that some of these features could arise from a dynamically active disk \citep{chequers2018bending} in isolation. 
Others note that ridges in phase-space may also connect to the Galactic bar \citep{muhlbauer2003kinematic, fragkoudi2019ridges}. 

Recent studies of the Orphan stream appear to challenge but also perhaps clarify much of this picture. It has been shown 
that stars in that stream have velocities that misalign with the stream track \citep{koposov2019piercing, fardal2019course}, 
and \citet{erkal2019total} have shown that it is possible to explain these offsets by the gravitational interaction with 
the LMC system if its mass (including an associated LMC DM halo) is 
$1.38\stackrel{+0.27}{{}_{-0.24}} \times 10^{11} M_\odot$, 
some 30 times more massive than its mass in stars 
\citep{vandermarel2006stars}
and 10 times 
more massive than an analysis of its rotation curve would suggest \citep{van2013third} 
--- though other authors have
also noted the need for a more massive LMC \citep{penarrubia2015timing, laporte2018response, 
moster2013abundance, Behroozi2013abundance} from very different viewpoints. 
\citet{erkal2019total}
have also used the motion of stars within the Orphan stream  to fit 
for the distorted shape of the matter DF in the MW from the 
interaction with the LMC.
In contrast to the distributions discussed earlier 
\citep{banerjee2011progressively, posti2019mass} --- the
shapes they determine are axially asymmetric, with some preference for a prolate geometry. 

We note the phase-space studies of \citet{antoja2018dynamically} were made from a sample of some 6 million stars. 
In this paper we consider a sample of up to 14.4 million stars, 
 carefully selected for sensitive studies of
the axial and vertical symmetry breaking patterns, 
to enable a determination of the most likely origin of the
observed effects. Remarkably, we find 
that the distorted matter DFs found by 
\citet{erkal2019total} and 
the asymmetries 
that we determine in our stellar data set can confront and discriminate between 
their offered solutions. 
In particular, 
we find strong preference for a prolate form, in loose agreement with 
earlier work \citep{helmi2004dark,banerjee2011progressively}, yet stemming from a
completely different origin. 

\section{Theory}

Noether's theorem reveals that 
each continuous symmetry of a Hamiltonian system 
has an associated integral of motion \citep{noether1918}.
In this paper we evaluate the extent to which 
MW stars out of the Galactic mid-plane region 
are axially symmetric,  
with the implication that 
axial-symmetry breaking would speak 
to the violation of the conditions under which $L_z$ holds  
as an integral of motion. 
Strictly speaking, this association requires that the 
converse of Noether's theorem holds (that, 
specifically, if $L_z$ is an integral of motion, then the system is
invariant under rotations about the 
$z$ axis). This holds here, noting Theorem 5.58 of \citet{olver2012applications}, with explicit 
demonstrations extant in the context of the stellar DF. 
For example, an isolated stellar system with an ergodic DF --- 
so that $f$ is a non-negative function of $H$ ---  is 
spherical \citep{binney2008GD}. Here non-zero, $L_z$ would imply that rotational symmetry
about the $\hat{z}$ direction should be manifest. Thus, 
if axial symmetry 
is broken, external,
and possibly time dependent, forces must be at work. 

In contrast, testing axial symmetry above and below the Galactic 
plane probes time-dependent interactions. That is, 
Theorem 6 of \citet{an2017reflection} states that an axially symmetric galaxy in steady state must have north-south reflection symmetry, where we note 
\citet{Schulz2013grav} for a slightly less general proof of north-south symmetry in steady state. 
Thus, a symmetry-breaking pattern in which axial symmetry is broken differently 
above and below the Galactic plane speaks to the existence of non-steady-state effects
within and possibly on the MW. 
To test axial symmetry, 
we count the number of stars on either side of $\phi = 180^{\circ}$, the anti-center line with 
$\phi$ the Galactocentric longitude, 
and compute the asymmetry parameter ${\cal A} (\phi)$:
\begin{equation}
    {\cal A} (\phi) = \frac{n_{L}(\phi) - n_{R} (\phi)}{n_{L}(\phi) + n_{R}(\phi)} \,,
    \label{AsymmetryParam}
\end{equation}
where $n_{L}(\phi)$ and $n_{R}(\phi)$ are defined as the number of stars 
at $|180^{\circ} - \phi|$, left and right of the anti-center line, 
respectively. 
The functions $n_{L,R} (\phi)$ 
subsume integrals over regions in the in-plane radial coordinate $R$ from the GC 
and the vertical distance $z$
from the mid-point of the Galactic plane. 
We note that Eq.~(\ref{AsymmetryParam}) implies that
$A(\phi)$ for the north plus south sample is not given by sum of the 
asymmetry in the north and that in the south. 
For a perfectly axially symmetric system, 
${\cal A}(\phi) = 0$. 

\subsection{External Torques from Nearby Masses} 

Our Galaxy possesses very 
massive satellite galaxies and is in the Local Group. The torques exerted by 
these external bodies could cause $L_z$ to be 
appreciably time dependent, spoiling axial symmetry. 
Non-steady-state forces could also exist within our sample, 
but in this section we consider torques stemming from forces external to it. 
In order to determine the most important contributions,
    we estimate the torques from the most massive and nearby objects beyond the MW, 
such as the Large and Small Magellanic Clouds (LMC and SMC) and the M31 (Andromeda) galaxy, 
as well as the Galactic bar, as it is not axially symmetric. 
We treat the Magellanic Clouds as a single system 
because they are bound together if the mass is in excess of 
$\sim 10^{11} M_\odot$ \citep{kallivayalil2013third}, and, moreover, 
its mass appears to be DM dominated, though we shall usually refer to this system as the LMC henceforth in this work.
We also evaluate the impacts of a 
few other prominent objects and show them to be 
relatively negligible. 
We assume the external sources are faraway point masses, ignoring the corrections that 
come from their finite extent. 
For the Galactic bar/bulge system we must be more careful. 
If the center of the MW
is co-located with 
the center of mass (CM) of the bar/bulge, 
symmetry 
constrains the torque from the CM to be zero. If its CM is at the mid-point of 
its length, its net dipole moment vanishes, yet 
it can still exert a non-zero torque because it has a small tilt 
with respect to the anti-center line. 
To compute the torques, we use the object locations  tabulated in SIMBAD \citep{wenger2000simbad}.
The Sun is taken to be at 
(-8,0,0) kpc in Galactocentric $xyz$-coordinates. 
The Galactic bar/bulge system consists of a box/peanut-shaped bulge and a long bar \citep{bland-hawthorne2016galaxy}, and we assume 
that the torque it exerts is dominated by the first, more massive object.  \citet{portail2015mass} have found its
dynamical mass to be $1.87 \pm 0.4 \times 10^{10} M_{\odot}$ within a box of 
$\pm 2.2 \times \pm 1.4 \times \pm 1.2\,{\rm kpc}$ 
in volume. The bar angle $\phi_{bp}$ made  
by its semi-major axis with respect to the anti-center line has been found from a study of 
red-clump giants to be $27\pm 2^\circ$ \citep{wegg2013mapping}, 
noting that the near side of that axis 
points in the first quadrant, $0^\circ < l <90^\circ$ \citep{wegg2013mapping, bland-hawthorne2016galaxy}. This crudely implies
that the half length of the peanut bulge is 1.4 kpc long, and for reason of estimate
we suppose a quarter of the dynamical mass is associated with 
the end of that half length. 
This gives our numerical value for the torque.
We compile these results in Table~\ref{tab:perturbers}, 
where $M$ is the mass of the external source, or 
perturber; $d$ is the distance of its CM from the Sun, 
which is the approximate center of our sample; 
and $\tau_z$ is the torque exerted by the CM on the Sun in the $\hat{z}$ direction. 
Our estimate for the Galactic bar is admittedly crude, but it should suffice for
our rough rank ordering. 

\begin{center}
\begin{table}[b]
    \centering
      \caption{\label{tab:perturbers} Nearby external objects 
      that torque the stars in our sample, with 
      torque reported in units of $M_{\odot}/$pc$^{-1}$. The errors in the inputs are such 
      that the LMC system undoubtedly gives the largest effect. Our torque computation does not use tidal forces only, but removing the difference would not change our assessment of the relative ranking of the perturbers. Note that the distant tide approximation \citep{binney2008GD} does not hold for all objects in Table \ref{tab:perturbers}. 
      }
    \hskip-2.45cm  
    \begin{tabular}{ |c|c|c|c|c|c| } 
       \hline
       Object & Mass ($M_{\odot}$) & distance (kpc) & $M/d^2$ ($M_{\odot}$pc$^{-2}2$)& $\tau_z$ ($M_{\odot}$pc$^{-1}$) \\ 
       \hline
       LMC (and SMC) & $1.4(3) \times 10^{11} $ \footnote{\citet{erkal2019total} \label{Erkal19}}& 52(2) \footnote{\citet{panagia1999distance} \label{panagia}} & 51 & 340,000  \\ 
       \hline
       M31 & $1.3(4) \times 10^{12}$ \footnote{\citet{penarrubia2015timing} \label{Penarrubia15}} & 772(44) \footnote{\citet{ribas2005first} \label{ribas05}} & 2 & -14,000  \\ 
       \hline
       Triangulum & $6 \times 10^{10}$ \footnote{Within 17 kpc from center as per \citet{corbelli2003dark} \label{corbelli2003}} & 839(28) \footnote{\citet{gieren2013araucaria} \label{gieren13}} &0.1 & -420  \\
       \hline
       Galactic Bar/bulge & $1.87(0.4) \times 10^{10}$ \footnote{\citet{portail2015mass} \label{robin12}} & 8 \footnote{Assumed \label{assumption}} & 288 & -47,000  \\ 
       \hline
       Sagittarius & $2.5(1.3) \times 10^8$ \footnote{\citet{law2010sagittarius} \label{Law2010}} & 28 \textsuperscript{\ref{Law2010}} & 0.3 & -240  \\
       \hline
       Fornax & $1.6(1) \times 10^8$ \footnote{\citet{lokas2009mass} \label{Lokas09}} & 138(8) \textsuperscript{\ref{Lokas09}} & 0.01 & 23  \\
       \hline
       Carina & $2.3(2) \times 10^7$ \textsuperscript{\ref{Lokas09}}& 101(5) \textsuperscript{\ref{Lokas09}}& $< 0.01$ & 16  \\
       \hline
       Sextans & $4.0(6) \times 10^7$ \textsuperscript{\ref{Lokas09}}& 86(4) \textsuperscript{\ref{Lokas09}} & 0.01 &  29  \\
       \hline
       Sculptor & $3.1(2) \times 10^7$ \textsuperscript{\ref{Lokas09}} & 79(4) \textsuperscript{\ref{Lokas09}} & 0.01 &  5  \\
       \hline
       \hline
       {\it Gaia}-Enceladus & ${\cal O}(10^9)$ \footnote{\citet{helmi2018merger,belokurov2018co} \label{helmi18}}& - & - & - \\
       \hline
   \end{tabular}
\end{table}
\end{center}

From Table~\ref{tab:perturbers}, it is apparent that the largest 
effect comes from the LMC system.
Other significant perturbers 
include the Galactic bar and M31, 
though the uncertainties are such that their 
relative roles could be reversed.
The net torque from these sources impacts 
both the shape and magnitude of ${\cal A}(\phi)$. Nevertheless our particular accounting 
shows that the LMC system grossly outweighs the other perturbers. However, if the shape of ${\cal A}(\phi)$ does not match that 
expected from the LMC, say, then this could speak to
matter effects, possibly from DM, that  
clandestinely torque our sample. 
Conversely, if we can account 
for the shape of ${\cal A}(\phi)$ we may well be able to constrain such structures. 
Thus far we have focused on 
external perturbations, which act to break the axial symmetry of 
our stellar sample, north plus south. However, non-steady-state
effects within our sample may also exist and stem from different
sources, such as from the passage of ancient satellites that 
perturb and excite the disk. Indeed, the interaction of 
the Sgr dwarf spheroidal (dSph) with the
Galactic disk has been 
suggested as the origin \citep{widrow2012galactoseismology,gomez2013vertical} of the 
vertical, wave-like perturbations we 
noted earlier \citep{widrow2012galactoseismology,yanny2013stellar,bennett2018vertical}, 
and the effect can also give
rise \citep{darling2019emergence,laporte2019footprints} to the {\it Gaia} phase-space spiral \citep{antoja2018dynamically}.

\section{Data Selection and Analysis} \label{sec:data}
We use data 
from the European Space Agency's {\it Gaia} space telescope, via the online {\it Gaia} archive \citep{prusti2016gaia, brown2018gaia}. 
The success of our analysis demands that we select stars, left and right, north and south, 
in a very balanced way.
Our selections were made from stars with measured parallaxes 
\citep{lindegren2018gaia}, though we choose to apply  
an intermediate offset of 0.07 mas (noting evidence for {\it Gaia} DR2 parallax zero-points
ranging from -0.029 to -0.083 mas depending on reference population in
\citet{zinn2019confirmation, stassun2018evidence, lindegren2018gaia}), 
to add to 
all parallax measurements.  
With the shift applied, 
we keep only stars with measured parallaxes, $\varpi > 0$ mas,  
though this shift is a trivial one for our
data set, because no stars are added as a result.
We also require $|b| > 30^{\circ}$ to avoid the extinction effects  characteristic of lower latitudes. To avoid selection bias, 
we remove all stars in the directions of the LMC and SMC,  
as well as their reflections across the mid-plane, across the anti-center line, and across both the mid-plane and anti-center line.  
The LMC and SMC are removed by requiring 
$b > -39$, $l \in [271,287]$ and $b \in [-41,-48]$, $l \in [299,307]$ respectively. 
The other six box cuts are constructed with 
suitable reflections.  
Considering 
the completeness of our data set in magnitude, color, $|z|$, and $R$
we see no clear evidence of incompleteness or of obvious, systematic biases if we choose 
$G_{\rm BP} - G_{\rm RP} \in [0.5, 2.5]$ mag, 
$G \in [14, 18]$ mag, $|z| \in [0.2, 3]$ kpc, 
$|b| > 30^{\circ}$, $\varpi > 0$ mas, and $R \in [7, 9]$ kpc. 
If we choose  $|180^{\circ}-\phi| < 12^\circ$,  
these cuts yield a sample of 14.4 million stars.  The key cuts which ensure completeness are restricting the {\it Gaia} data sample to brighter limits ($G < 18$) and avoiding crowded low latitude regions. Tests involving restrictions to an even brighter limit $G<17$, while lowering significance with a smaller sample, does not change our asymmetry findings (see Figure 1) and gives us confidence that we are not probing incomplete {\it Gaia} DR2 samples as a function of azimuth.
We defer more discussion of the completeness studies
that motivate these choices to a future paper (Hinkel et al., 2019, in preparation),  
though we find it pertinent to highlight a key result of that 
work: as a result of our selections in $G$-band magnitude
and color, we find the average relative parallax error of our stars to 
be reduced to some 10\%, even though we have not directly restricted
that parameter, because the distance distributions would become
skewed as a result \citep{luri2018gaia,bailer-jones2018distance}. 
Moreover, noting Fig.~7 of \citet{arenou2018gaia}, we have also
explicitly studied the impact of more crowded fields on our
results.
We find, e.g., that making additional restrictions 
on our data set in the direction of the GC has a negligible impact on
the results we report here. 

Table ~\ref{tab:raw counts} shows that our data selections are
well matched, north and south, as well as left and right, 
showing no sign that 
spatial asymmetries in the dust observed in the mid-plane region \citep{schlegel1998dust} 
impact our results. 
The left and right samples, north and south, are matched to within about 0.06\%.  
The larger, but still very small, number count asymmetries we observe in the north or 
south turn out to match more poorly, 
but its source may stem from the 
physics that makes $A_{{\rm N,S}}(\phi)$ so much larger. 
As the $\phi$ dependence of $A(\phi)$ is our key result, we have also 
studied completeness within the $x-y$ plane carefully to determine 
that we should limit $|180 - \phi| \le 6^\circ$ for our 
$R$ selection, implying, roughly, 
that we choose a reach in $x$ and in $y$ that is about $\pm 1\,{\rm kpc}$ of the
Sun's location, yielding a sample of 11.7 million stars. 

\begin{center}
\begin{table}[ht!]
    \centering
    \caption{\label{tab:raw counts}The number of stars found in each quadrant of the analysis, with $|180^{\circ} - \phi| < 12^\circ$. 
    Totals for the left and right are also shown.  The sample is very
    evenly distributed, 
    left and right, with 
    an aggregate asymmetry of ${\cal A} \approx 6 \times 10^{-4}$. 
    }
    \hskip-2.45cm  
    \begin{tabular}{ |c|c|c|c| } 
       \hline
       \ & Left & Right & 
       Asymmetry (\%) \\ 
       \hline
       North & 3,376,969 & 3,471,980 & -1.39\\ 
       \hline
       South & 3,815,477 & 3,729,647 & 1.14\\ 
       \hline
       TOTAL: & 7,192,446 & 7,201,627 & -0.06\\ 
       \hline
   \end{tabular}
\end{table}
\end{center}

\subsection{Data Analysis} \label{Sec: Data Anaylsis}

\begin{figure}[ht!]
\begin{center}
  \subfloat[]{\includegraphics[scale=0.59]{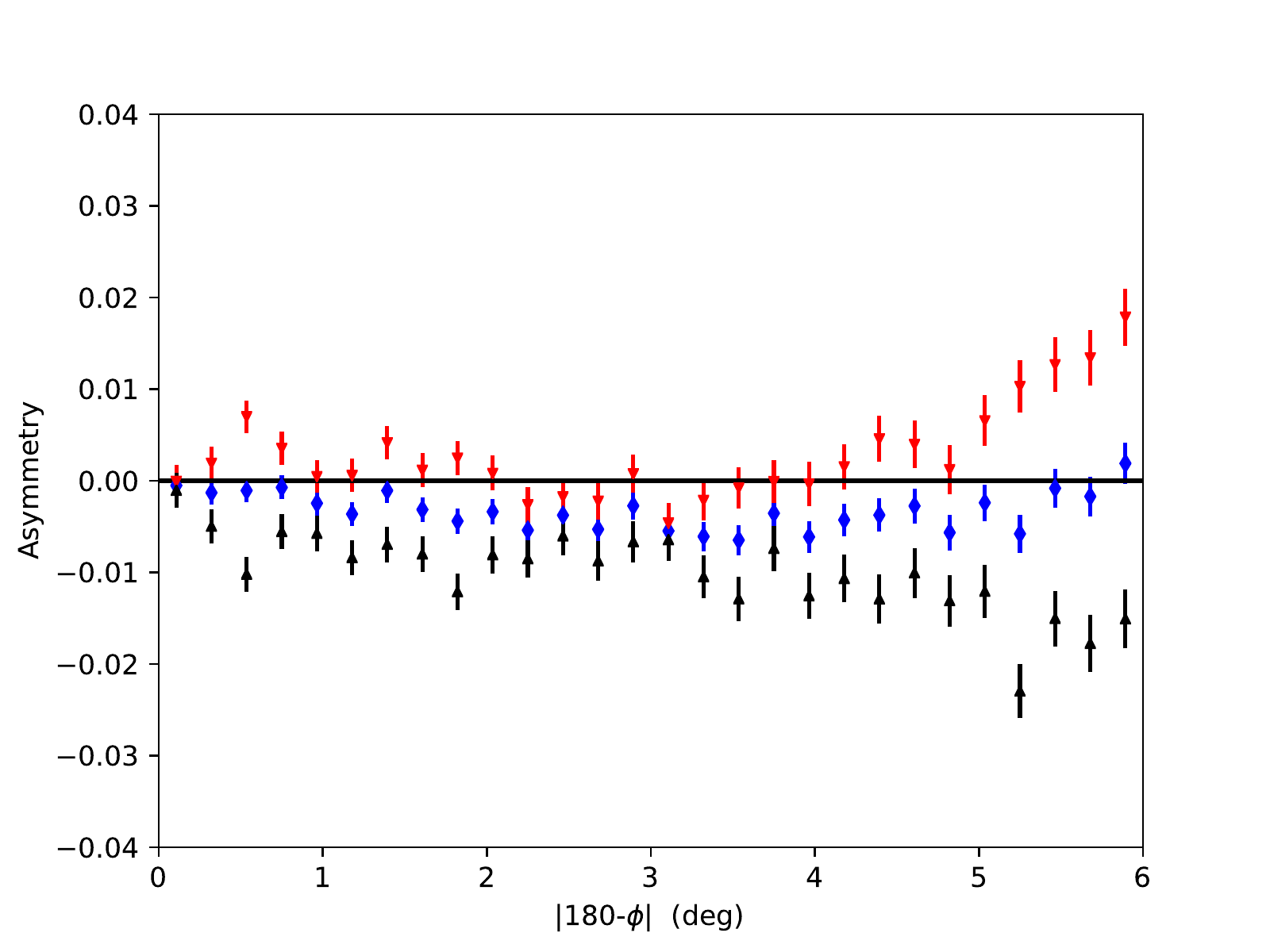}}
  \subfloat[]{\includegraphics[scale=0.59]{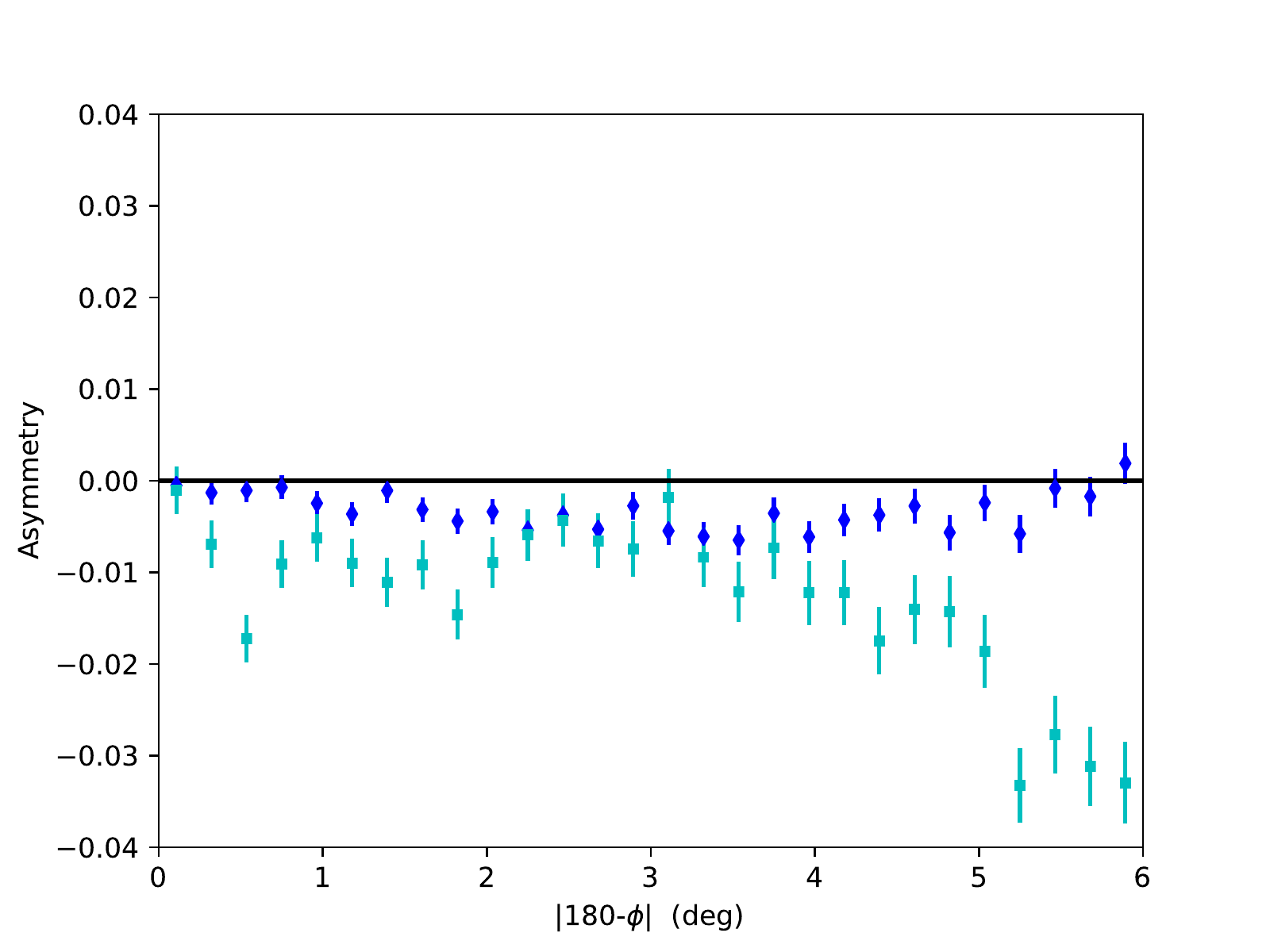}}
  
\subfloat[]{\includegraphics[scale=0.59]{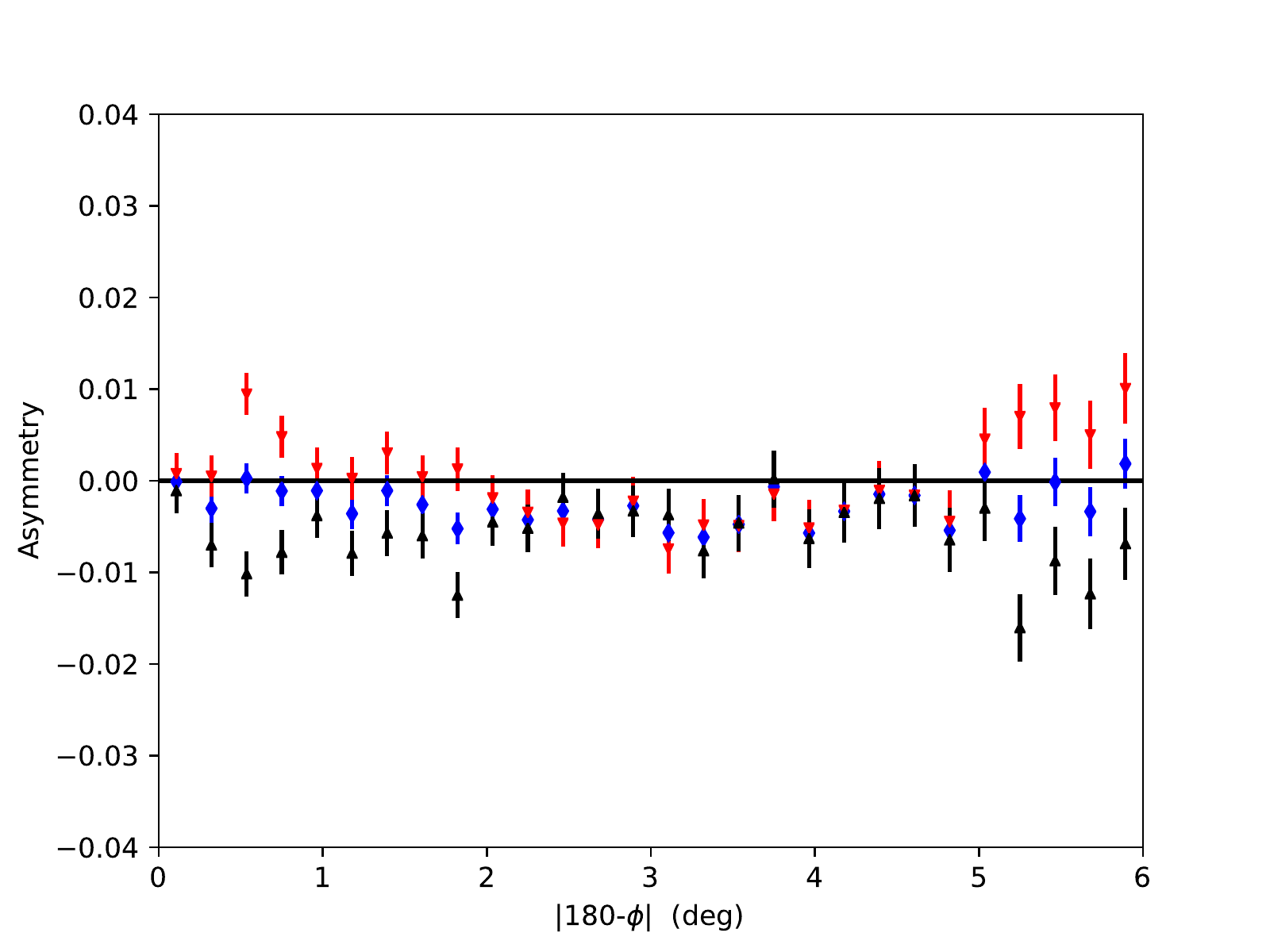}}
\subfloat[]{\includegraphics[scale=0.59]{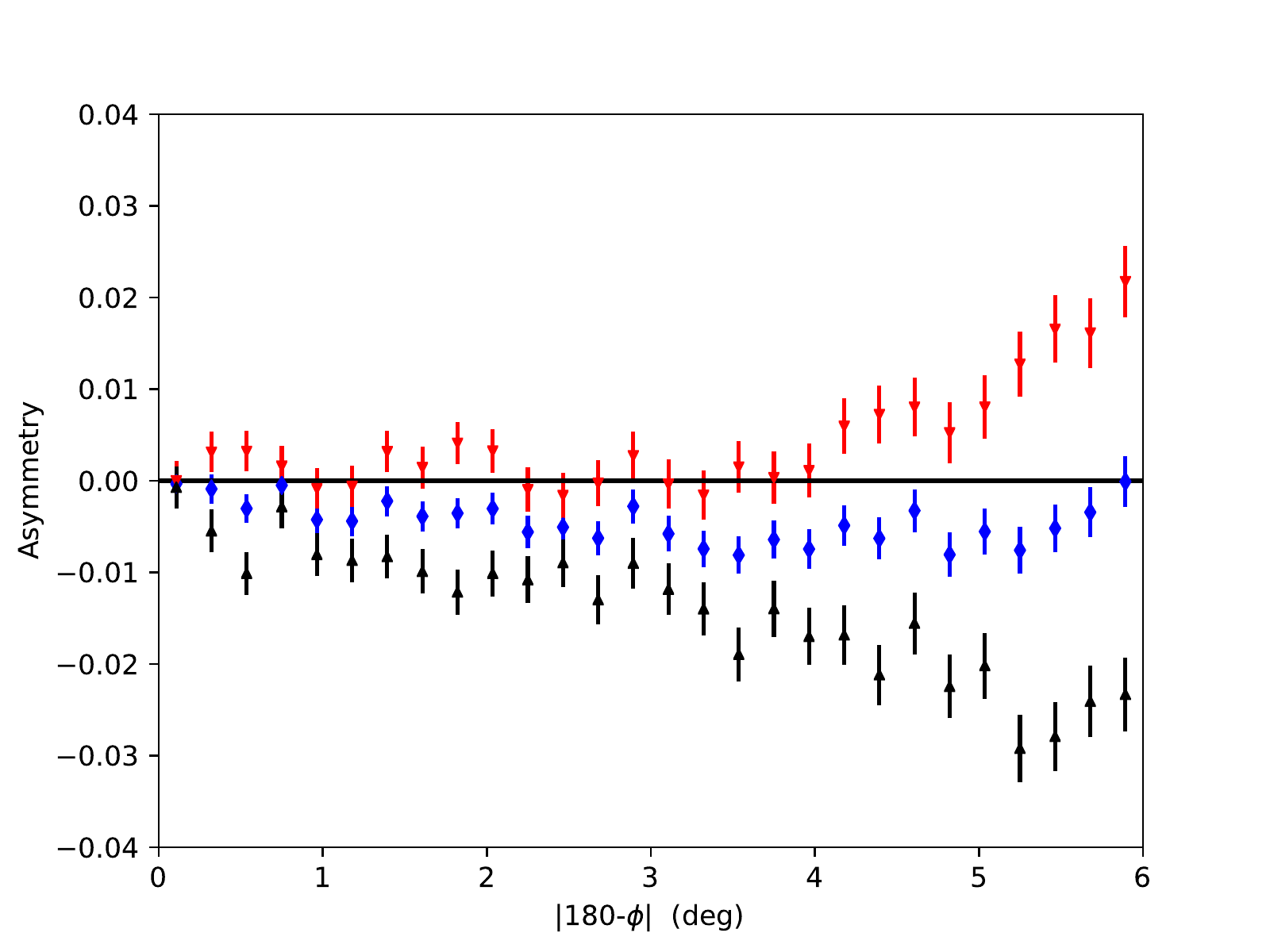}}
\caption{
Asymmetry $A(\phi)$ with $\phi$ for
(a) 
our selected 
data set, with red, downward pointing triangles (S);  
black, upward pointing triangles (N); and blue diamonds (N+S). (b) We compare $A(\phi)$
in the  N+S sample with the difference of $A(\phi)$ 
in the north and $A(\phi)$ in the south (N-S; squares). 
We compare these results
with different $G$-band magnitude selections, in (c) $16 < G < 18$ mag, noting that  
by doubling the size of our 
magnitude window, we do not appreciably change our result, and (d) $14 < G < 17$ mag, 
minding \citep{luri2018gaia}, where we note the text for further discussion. 
Here, too, there is no significant, 
qualitative change when we include stars with $G<18$. 
}
\label{fig:axialwithG}
\end{center}
\end{figure}

The results of our asymmetry analysis of star counts left-right of the anti-center line are shown in 
Fig.~\ref{fig:axialwithG}, with panel a) 
revealing that 
axial symmetry 
in the north plus south (N+S) sample (blue diamonds)
is significantly broken at a level up to $0.5\%$ out to angles $|180-\phi| < 6^{\circ}$,
though the symmetry breaking effects in the north (N) only (black up triangles) or south (S) only (red down triangles)
samples can be much larger. 
Remarkably the N and S 
left-right asymmetries are also anti-correlated in sign, so that 
the difference in the N and S asymmetries can be grossly larger than 
that of the N+S sample as shown in Fig.~\ref{fig:axialwithG}b. This comparison shows that  
the symmetry-breaking effects from 
non-steady-state interactions within and beyond the Galaxy are grossly 
larger 
than those resulting from a steady, external perturbation. In panels c) and d) we reinforce the results of panel a) by noting that the asymmetry trend persists when only keeping stars with $16 < G < 18$ (panel c) and when making a very conservative faint end cut, keeping stars with $14 < G < 17$ (panel d).
\citet{luri2018gaia} note that the {\it Gaia} DR2 catalog is ``essentially 
complete between $G\approx 12$ and $\sim \!17$ mag,'' though it also extends 
significantly 
beyond $G=20$ mag. Parallax measurements are, however, quite incomplete for $G > 18$ in {\it Gaia} DR2, and \citet{luri2018gaia} also remark that the faint end limit is ``fuzzy'' in that it can depend on object density and on the filtering on data quality prior
to publication. Nevertheless, we do not observe any significant 
changes in our results with 
changes in the {\it G}-band selection so long as we 
choose $G < 18$ magnitude cuts. Thus we opt for the largest selection we can make.  If we restrict to a brighter limit than $G<17$, then substantially decreased number counts do more strongly begin to compromise the significance of especially 
N (only) and S (only) studies of the asymmetry.

\begin{figure}[ht!]
\begin{center}
    \subfloat[]{\includegraphics[scale=0.35]{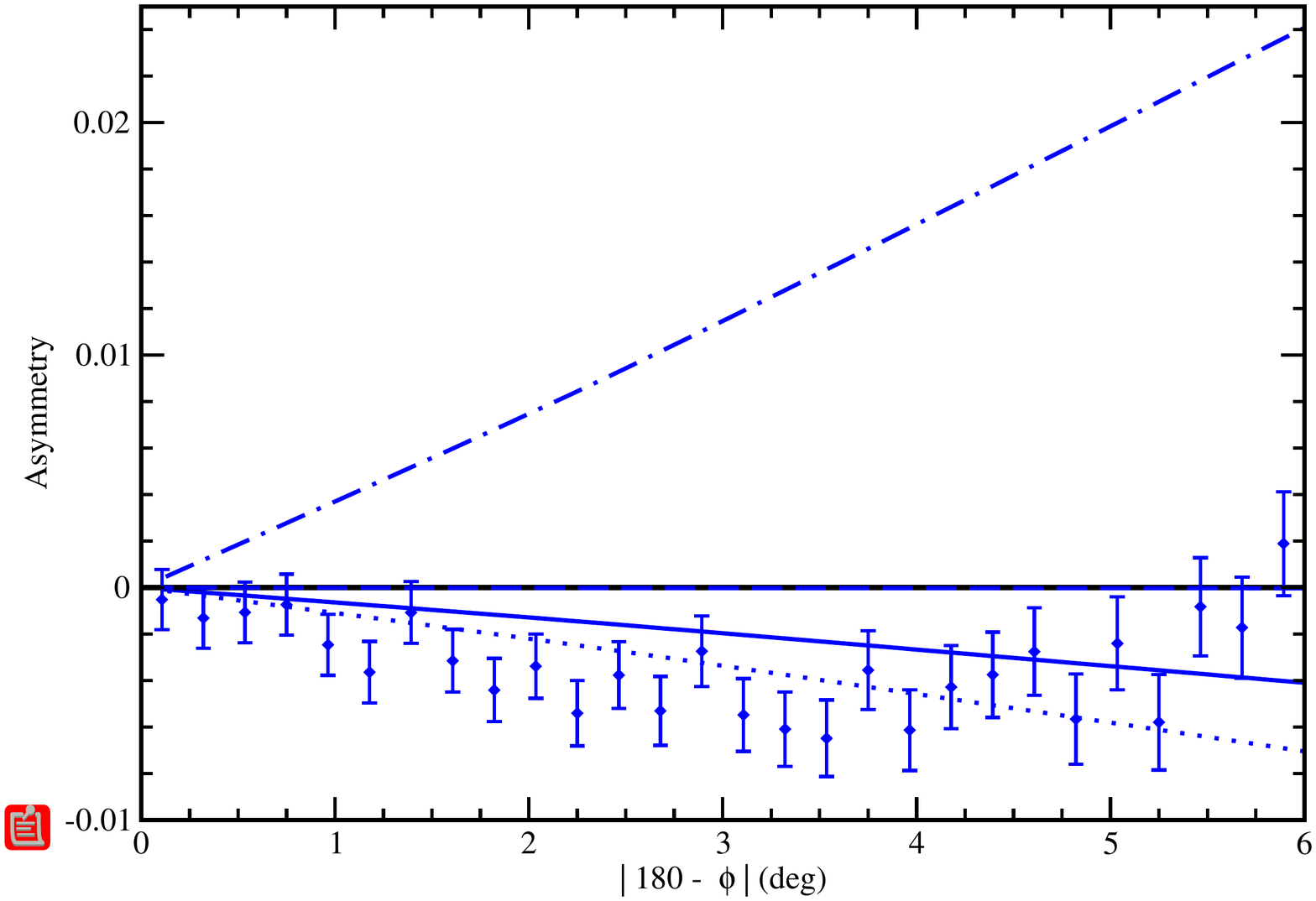}}
    \subfloat[]{\includegraphics[scale=0.35]{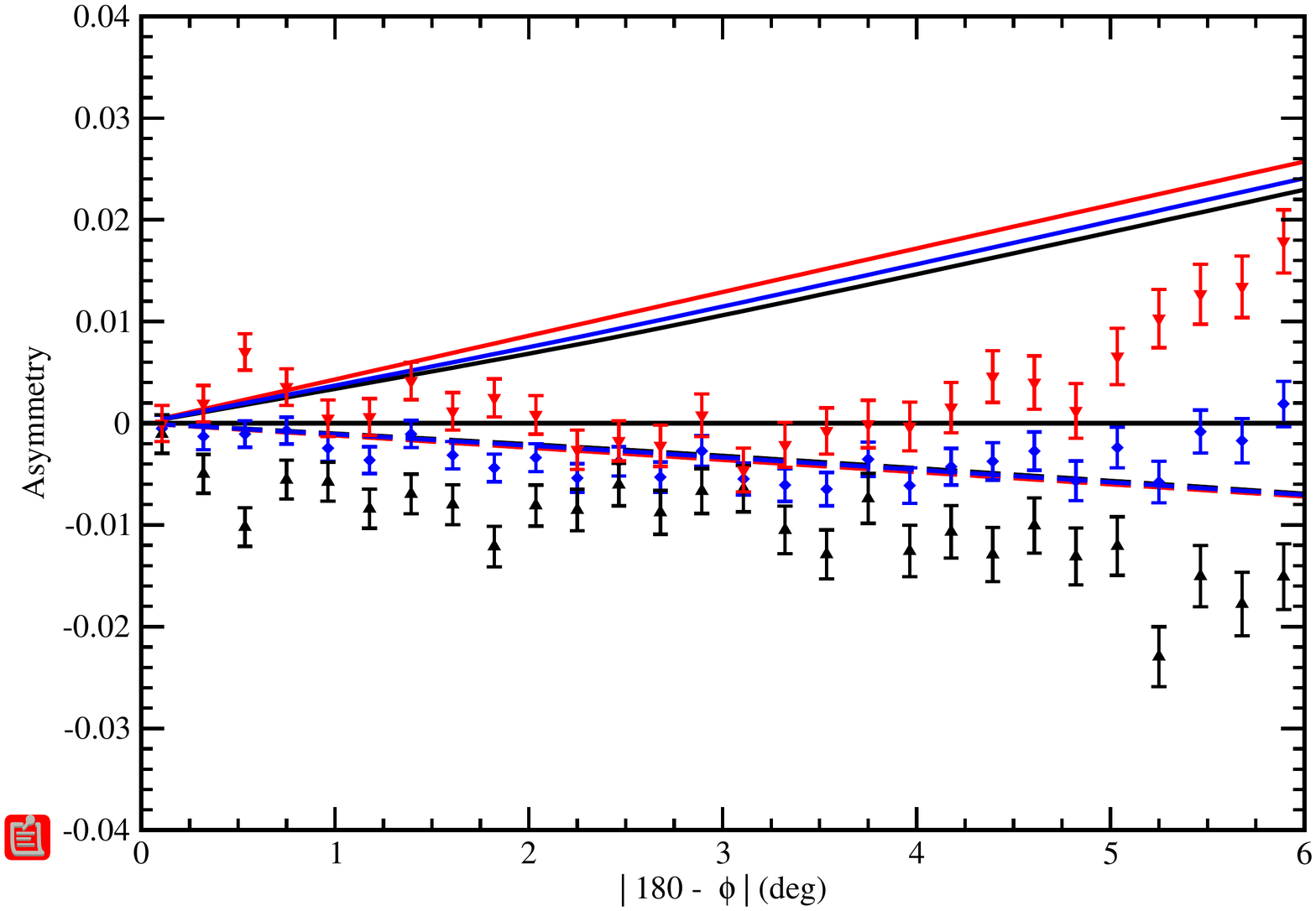}}

    \subfloat[]{\includegraphics[scale=0.35]{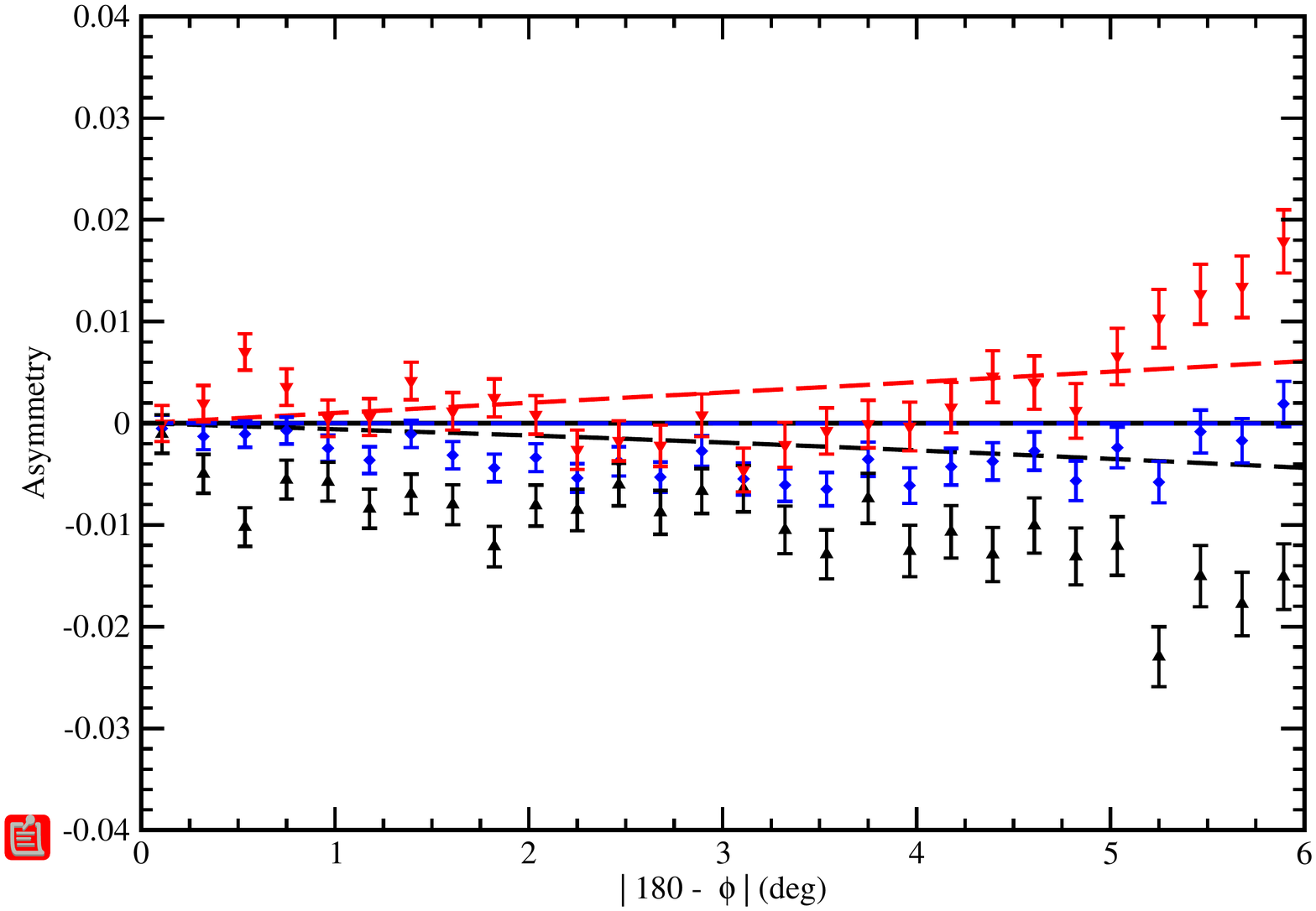}}
    \subfloat[]{\includegraphics[scale=0.35]{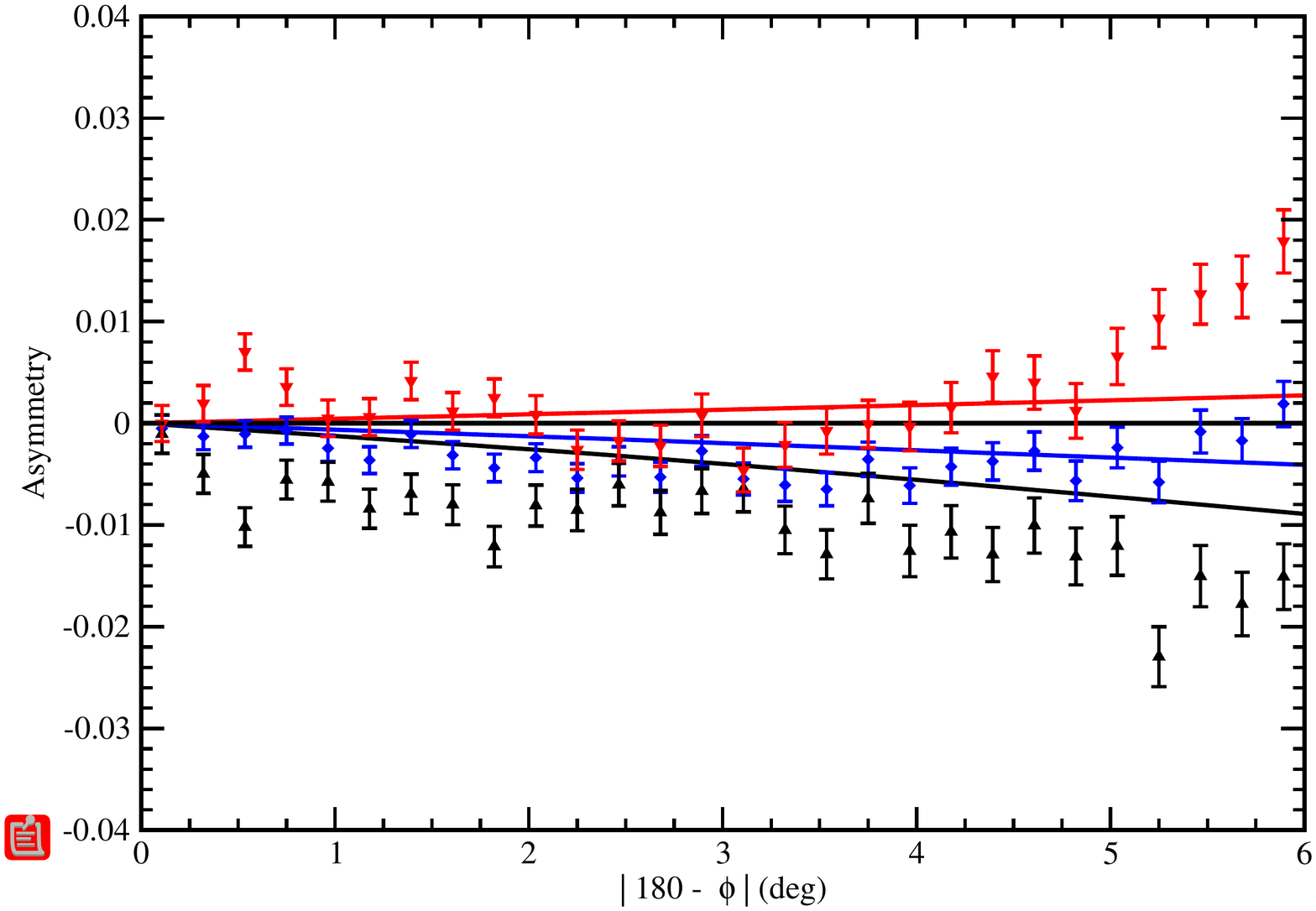}}
 \caption{
Asymmetries, as in Eq.~(1), computed for the geometry of our sample 
using the distorted MW halo models of \citet{erkal2019total} 
from fits of the LMC on the Orphan stream, with and without the reflex motion of the
MW, are compared with the results of Fig.~\ref{fig:axialwithG}a). 
In (a) we show the N+S asymmetry of 
Fig.~\ref{fig:axialwithG} with the oblate (dotted), reflex oblate (dotted-dashed), and
reflex prolate (solid). The prolate result has also been included, but its asymmetry is so 
small that it is indistinguishable from the horizontal axis. In (b) we compare the
asymmetries from Fig.~\ref{fig:axialwithG}a with those for 
the oblate (dash) and reflex oblate (solid), 
for S (red), N (black), and N+S (blue), and use these latter assignments throughout. 
We compare with the prolate (dash) results in (c) and the reflex prolate (solid) 
results in (d). 
}   
\label{fig:allAxial}
\end{center}
\end{figure}

Although the asymmetries we have found are small, they are nevertheless significant. 
For the N+S result shown in Fig.~\ref{fig:axialwithG}a, a linear fit to the 
data shows that both the constant and linear term are non-zero beyond $5\sigma$ 
significance: 
$A(\phi)= (-2.0963 \pm 0.0003 + |180 -\phi|(-0.45050 \pm 0.00003))\times 10^{-3}$. 
The N+S asymmetry is larger than 
the aggregate raw number count asymmetry of our data selection, 
shown in Table~\ref{tab:raw counts}. Were we to repeat 
the raw number count comparison
for a maximum value of $|180 -\phi|=6^\circ$ 
we would find a value of -0.0032, 
so that our fit result is also significantly different from that. 
It is thus apparent 
that none of the asymmetries --- N, S, or N+S --- are constant 
with $\phi$. Moreover, 
 an anti-correlation of the asymmetries
N and S is also present, noting that at values of 
$|180^{\circ} -\phi| \sim 0.5^\circ, 1.8^\circ,$ and $ >5^\circ$ (Fig. \ref{fig:axialwithG}b),
an increase
in the asymmetry in the N is matched by a more negative asymmetry in the $S$. We comment on these smaller scale asymmetries briefly below.

\subsection{
Asymmetries from Mass Distribution Models Deduced from Orphan Stream Fits} \label{Sec: Model Anaylsis}

\citet{erkal2019total} have computed the relative likelihoods 
of several different MW mass distribution models 
that were used to fit out-of-stream velocities in 
the entire
Orphan stream in the presence of the LMC.
They adopt a generalized form of  
``MWPotential2014'' from \citet{bovy2015galpy}, which 
consists of bulge, disk, and halo components, though they keep the bulge and disk
components fixed and  allow only the mass and shape of the halo 
to vary.
\citet{erkal2019total} find that the LMC 
induces a flattening of the halo in a direction away from the $z$ axis, though they 
caution against interpreting this as an intrinsic
property of the disk; we revisit this in the face of our 
asymmetry results below.
The distortion of the halo, which they assume is of NFW form \citep{navarro1997universal},  
 yields either an oblate or prolate shape, and they allow
for the reflex motion of the MW in the presence of the LMC, though they have not 
allowed the shape of the LMC itself to vary. 
(For reference, we note that the scale height in their 
disk model is 280 pc \citep{miyamoto1975disk}.) 
We have taken the various best fit parameters from Table A1 
of  \citet{erkal2019total} and have computed an additional observable: 
the asymmetry in the left-right star counts versus Galactocentric azimuth 
(in N, S, and N+S versions) that would result 
from each of the distorted halo 
models that they tabulate 
in their Table A1 ---  oblate, prolate, and with and without the reflex action of
 the MW halo. We have not used their spherical halo solution, which their fits strongly 
disfavor, because that would yield a vanishing left-right asymmetry. 
That we expect these asymmetries to be pertinent to our asymmetry results stem from the 
torques estimated in Table ~\ref{tab:perturbers} --- it is apparent that the LMC 
must dominate any 
 N+S asymmetry today. Moreover, although 
\citep{erkal2019total} fit for a distorted halo only, the 
outcome of their work is a distorted matter distribution, 
which we probe near the Sun 
through the distribution of stellar number counts.
The results of these analytical computations are shown in straight lines in Figure 2, 
overlaid with the star-count results from Figure 1.

The data in panel a) of 
Fig.~\ref{fig:allAxial} clearly show 
a significant large scale trend in the left-right 
asymmetry of combined N+S star counts (blue diamonds) 
extending from the $\phi = 180^{\circ}$ anti-center line to $\sim 6^{\circ}$ left or right.  
Evaluating the $\chi^2$ statistic for these computed models compared with the data, noting that the 
\citet{erkal2019total} fits contain
five parameters: the halo mass and scale radius, and the magnitude and orientation 
(in $(l,b)$) of the flattening, we find  for our 28 data points
that 
$\chi^2/(n=23)$ is 3.64 (reflex prolate, solid blue with a downward tilt), 6.40 (oblate, dotted with downward tilt), 7.11 (prolate, solid, hugging $A\sim 0$), and 
201 (reflex oblate, dotted-dashed, upward tilt).
Although no one model describes the N+S asymmetry data well, 
the preference for the reflex prolate or possibly the oblate (though see below) description is clear. 
That is, of the models considered by \citet{erkal2019total}, 
an LMC torque which distorts the halo of the 
MW, while accounting for its reflex motion, 
into a prolate ellipsoid with its major axis aligned roughly 
along the line between the GC 
and the LMC provides the best match to our solar-neighborhood star-count asymmetries. 
An oblate model that is not corrected for reflex motion can 
also fit the combined $N+S$ data, but once we consider the N and S
samples separately the oblate halo models are clearly ruled out.   
All the models are more strongly distinguished once the N and S results are also considered, and 
we show these comparisons in panels b), c), and d) of Fig.~\ref{fig:allAxial}. 
There we compare our asymmetry count results with those of the model fits in the N, S, and N+S, 
where we consider oblate (dashed) and reflex oblate (solid) in b), prolate in c)
and reflex prolate in d). The problem with matching the data to an oblate model 
becomes apparent when looking at the N, S, and N+S asymmetries in 
Fig. \ref{fig:allAxial} b): the three curves are nearly coincident, but the order of the lines is reversed, with S slightly  below N in the model calculations, whereas the N-only asymmetry is much more negative than the S-only in the data counts,
in clear contradiction with the data.  Additionally, the split in N-only and S-only tracks is not reflected in the nearly coincident model 
lines. 
Comparing with the reflex oblate model, we see that the N, S, and N+S asymmetries
split slightly, but they all very much diverge from the data.  
In panels c) and d), we see that the prolate and reflex prolate models have asymmetries that 
strongly differentiate
N and S, as do the data points from Fig. \ref{fig:axialwithG}. The prolate model has a near null N+S asymmetry; this results because its 
major axis points very nearly in the $l=90^\circ$ direction, so that in summing 
N and S there is no left-right asymmetry. We thus see that the reflex prolate model describes
the data better. These conclusions are very much born out by 
a $\chi^2$ analysis; for N, S, and N+S, respectively, 
$\chi^2/(n=23)$ is 15.8, 21.2, and 6.40 (oblate); 
143, 88.7, 201 (reflex oblate); 
18.7, 12.2, 7.11 (prolate), and 
15.0, 9.82, 3.64 (reflex prolate). 
We thus conclude that a oblate shape in which the flattened direction is in the orbital plane 
of the LMC, needed to fit the Orphan stream data \citep{erkal2019total} is ruled out. 
Thus, by showing that a prolate, 
reflex halo model is best fit (amongst the small set of models here) and by ruling out 
oblate models, we demonstrate the power of asymmetries to make new and significant 
constraints on the distribution of DM in and around our MW.
%
%
%
%

While tying this overall $\pm 6^{\circ}$ trend in $\phi$ to the influence of a massive LMC 
and demonstrating its influence on the DM halo of our MW, distorting it 
into a prolate spheroid, is our main result, we also note several smaller scale ``blips'' 
in the asymmetries of Fig.~\ref{fig:allAxial}b) 
which may be attributed to some of the other substructures listed in 
Table ~\ref{tab:perturbers}.  We discuss this 
further in \S ~\ref{subsec:NonSteadyState}.

\section{Results} 

\subsection{Evidence for External Perturbations}
\label{subsec:NonIsolationEffects}

The LMC appears to be the dominant external
influence on 
the Galaxy.  \citet{erkal2019total} find a galactic potential that incorporates the 
LMC (and SMC)'s effect on the MW, 
and we note that it explicitly breaks axial symmetry. 
Upon integrating their models over the same volume of 
space as that used in Fig.~\ref{fig:allAxial} a), we find that the ``reflex prolate'' 
model of \citet{erkal2019total} is the most consistent with the observed axial asymmetries.  

While detailed model explanations are beyond the scope of this work, we note three 
further possible connections between the LMC and 
non-axisymmetric structure in the disk and halo, which have already been 
suggested in the literature:

1. As first pointed out in \citet{law2010sagittarius}, the pole of the Magellanic stream is aligned within $1^{\circ}$ of the tilted triaxial MW halo needed to reproduce the orbit of the Sgr stellar stream. Increasing the mass by a factor of a few, 
as suggested by \citet{erkal2019total}, makes the apparent unusual alignment and shape of the MW halo compared with its disk quite plausible.

2. An analysis by \citet{iorio2018shape} of the distribution of RR Lyrae associated with the
{\it Gaia}-Enceladus structure \citep{helmi2018merger,belokurov2018co} have been shown to point to a MW halo elongated in the direction to the LMC. 

3. The line of anti-nodes for bending modes in the HI gas disk \citep{levine2006vertical} is at $\phi\sim 270^{\circ}$ and for the Cepheid-traced outer stellar disk \citep{skowron2019three, chen2019intuitive} is in the range $\phi \in [245^{\circ}, 255^{\circ}]$, not far from $\phi_{\rm LMC}\sim 269^{\circ}$.  The orientation of the long axis of the prolate halo geometry we favor coincides with this direction as well and could support this $m=1$ bending mode as suggested by \citet{dekel1983galactic}, \citet{sparke1988model}, and \citet{ideta2000time}, 
helping to explain its long-lived nature \citep{ideta2000time}.
Linking the LMC to the warp also supports the results of \citet{weinberg2006magellanic}. 

\subsection{Evidence for Non-steady-state Effects} \label{subsec:NonSteadyState}

We argue that the largest MW perturber is a heavy LMC system having 
some 10\%
of the MW's mass \citep{erkal2019total}.  That system, 
assumed to be on first passage by the MW, 
has a typical median infall 
time of $\sim$1.4 Gyr for 
MW and LMC masses similar to what 
we assume here \citep{patel2016orbits}. This timescale is 
long enough to be considered quasi-steady-state and 
results in unobservably slow adiabatic changes \citep{binney2008GD}. 
Nevertheless, we regard the observed distortions not as long-term
properties of the disk, keeping in mind that 
it is difficult to realize a stellar disk
that is misaligned with the halo 
\citep{debattista2013orientation, erkal2019total}, but rather as a response
to the LMC infall. 
Yet, there are 
regions where the left-right asymmetry 
is much larger, particularly if we consider the asymmetry N-S 
rather than  N+S, 
as in Fig.~\ref{fig:axialwithG}b) near  
$|180^{\circ}- \phi| \sim 0.5^\circ, 1.8^\circ, >5^\circ$.  
According to Theorem 6 from \citet{an2017reflection}, the approximate azimuthal symmetry 
here means than the north-south difference we see is indicative of a departure from 
steady-state dynamics on smaller time --- and length --- scales. 

Given that the effect that causes these ``blips'' 
should be appreciably time dependent, 
the Galactic bar is a great candidate, 
with a pattern speed known to be roughly $39 \pm 3.5 \ \sfrac{ \rm km}{ \rm s \cdot kpc}$ \citep{portail2016dynamical}.  
This pattern speed corresponds to a period of roughly 160 Myr, much shorter than the dynamical 
timescale of the LMC infall 
and is comparable with the crossing time near the solar neighborhood ($\sim 300$ Myr). 
This hypothesis is bolstered by the fact that the Outer Lindblad Resonance is 
thought to be near the solar circle \citep{dehnen2000effect}, 
where we also
note \citet{fragkoudi2019ridges}. 

Generally, the emergence of features that differentiate N from S 
supports our interpretation of the halo distortion, 
which we also observe through axial asymmetries in our stellar sample, as a 
response to the LMC infall.

\section{Summary}

We have discovered statistically significant left-right and 
north-south asymmetries in {\it Gaia} DR2 star counts 
in the solar neighborhood, which are all consistent 
with a large scale perturbation caused primarily by the LMC system --- and its associated 
DM.  Previous discussions of the relative influence of the 
LMC on 
MW 
disk asymmetries \citep{hunter1969dynamics} 
would underestimate the LMC's relative influence 
due to early, lower-mass estimates of the LMC 
and tidal force approximations, which may not work well 
when the larger LMC/MW mass ratio of some 10\% is used. 
Now, recent work by \citet{erkal2019total}, 
with its significantly larger
and more accurate LMC mass, 
gives significant credence to the suggestion by \citet{weinberg2006magellanic} 
that the LMC could in fact be nearly entirely responsible for 
the long observed HI gas warp of the MW disk.
Moreover, when then modeling 
the LMC's influence, a non-reflex model
which assumes $\rm M_{\rm LMC} << \rm M_{\rm MW}$, is insufficient, 
and one that can include reflex reactions of the MW due to the LMC, such as that in  \citet{erkal2019total}, is more appropriate. 
We find now, that 
not only can the LMC's influence explain 
the HI gas warp, 
but it also appears to induce a substantial asymmetry in the star counts left 
vs. right and north vs. south in the solar 
neighborhood of the correct sign and magnitude. 
Looking at other possible perturbers, 
the effect of the LMC is dominant compared to that of 
the Galactic bar (in most scenarios), 
the Sgr dwarf and stream, 
and finally also 
the more massive but much more distant 
perturbers, such as M31 (see Table 1). The odd, tipped 
triaxial shape of the MW's dark halo suggested by \citet{law2010sagittarius} based on the 
orbit of the Sgr stream 
and the elongation of the {\it Gaia}-Enceladus structure \citep{helmi2018merger,belokurov2018co} 
can also both be potentially more simply understood by 
the gravitational interaction with 
the LMC
--- though detailed modeling remains to confirm 
that these additional suggestions 
do operate in detail.

Deviations from symmetry 
in the case of star counts near the sun (at only the sub-percent level), 
combined with results related to Noether's theorem associating a conserved angular momentum with rotational 
symmetry 
are shown here to be powerful probes of the influence of satellite torques on the overall distribution of mass in and around the MW.

\acknowledgments{
S.G. and A.H. acknowledge partial support from the U.S. Department of Energy under
contract DE-FG02-96ER40989. We thank Scott Tremaine for bringing \citet{an2017reflection} to 
our attention and Issac Shlosman for a discussion of Galactic warps. S.G.
thanks The Galileo Galileo Institute for Theoretical Physics for hospitality 
during the completion of this work. 
A.H. thanks the Universities Research Association and the GAANN fellowship for funding.
We thank the anonymous referee for thoughtful comments that improved the paper. 

This document was prepared in part using the resources of the Fermi National Accelerator
Laboratory (Fermilab), a U.S. Department of Energy, Office of Science, HEP User Facility.
Fermilab is managed by Fermi Research Alliance, LLC (FRA), acting under Contract No.
DE-AC02-07CH11359.

This work has made< use of data from the European Space Agency (ESA) mission
{\it Gaia} (\url{https://www.cosmos.esa.int/gaia}), processed by the {\it Gaia}
Data Processing and Analysis Consortium (DPAC,
\url{https://www.cosmos.esa.int/web/gaia/dpac/consortium}). Funding for the DPAC
has been provided by national institutions, in particular the institutions
participating in the {\it Gaia} Multilateral Agreement.

}

\bibliography{main_GHY_arxiv_9dec2019v2.bbl}

\end{document}